\documentclass[aps,prl,twocolumn,showpacs,letterpaper]{revtex4-1}
\usepackage[charter]{mathdesign}
\usepackage{amsmath}
\usepackage[pdftex]{graphicx}
\usepackage{microtype}
\usepackage{bm}\let\vec\bm
\usepackage[svgnames]{xcolor}
\usepackage[colorlinks=True,linkcolor=DarkRed,citecolor=ForestGreen,urlcolor=MediumBlue,pdfstartview=FitH,bookmarks=False,pdfpagemode=UseNone]{hyperref}
\begin{document}
\title{Causality, Non-Locality and Negative Refraction}
\author{Davide Forcella}
\email{davide.forcella@espci.fr}
\author{Claire Prada}
\email{claire.prada@espci.fr}
\author{R\'emi Carminati}
\email{remi.carminati@espci.fr}
\affiliation{ESPCI Paris, PSL Research University, CNRS, Institut Langevin, 1 rue Jussieu, F-75005, Paris, France}
\begin{abstract}
The importance of spatial non-locality in the description of negative refraction in electromagnetic materials has been put forward recently. 
We develop a theory of negative refraction in homogeneous and isotropic media, based on first principles, and that includes non locality in 
its full generality. The theory shows that both dissipation and spatial non locality are necessary conditions for the existence of negative refraction.
It also provides a sufficient condition in materials with weak spatial non locality. 
These fundamental results should have broad implications in the theoretical and practical analyses of negative refraction of electromagnetic
and other kinds of waves.
\end{abstract}
\maketitle

%%%%\section{Introduction}

The study and design of artificial materials (metamaterials) with exotic electromagnetic properties has attracted a lot of attention both at the theoretical and experimental levels. 
Negative refraction has remained one of the most intriguing properties: in a certain range of frequencies, the energy flow is opposite to the direction 
of the phase velocity~\cite{veselago1968,smith2000,pendry2000,shelby2001,VeselagoReview}, offering a new potential for the design of components 
and devices (the flat lens being the most famous example).
The investigations of the conditions for negative refraction have been originally carried out using the local description of electrodynamics 
in continuous media in terms of the electric permittivity (dielectric function) $\epsilon(\omega )$ and magnetic permeability $\mu(\omega )$ (for a review see for example Ref.~\cite{VeselagoReview}). 
The first metamaterials exhibiting negative refraction were severely limited by absorption, and various strategies have been explored to design almost 
transparent media. Nevertheless, basic principles of electromagnetic waves propagation
impose general constraints on the response functions of materials, and on the conditions for negative refraction. 
For example in Ref.~\cite{stockman2006} the principle of causality has been used 
to argue that, for spatially local materials, dissipation is necessary for the existence of negative refraction.
More recently, it has been pointed out that the use of local electrodynamics to analyse negative refraction 
in certain artificial materials has some intrinsic flaw. The local effective $\epsilon(\omega )$ and $\mu (\omega )$, usually obtained by fitting
reflectivity or transmissivity curves, exhibit physical inconsistencies. The necessity to include spatial non-local corrections has been 
put forward, in order to overcome the inconsistencies in the recovered parameters~\cite{chebykin2012,grigoriev2014,belov2003, orlov2011, chebykin2011,Alu2011}. Independently, it has been shown that spatial non-locality is the fundamental 
ingredient for the existence of negative refraction in some specific materials (including natural materials)~\cite{forcella2014,amariti2011}. 

Although the role of spatial non locality in the phenomenon of negative refraction has been highlighted in recent studies, a theory 
providing a general understanding of the conditions for negative refraction, based on first principles and including spatial non locality, is still missing.
The purpose of this Letter is to develop such a theory, and to analyse its consequences in terms of necessary and sufficient conditions for the
observation of negative refraction in natural and artificial electromagnetic materials. The analysis leads to the intriguing conclusion that 
negative refraction for electromagnetic waves is not possible in absence of spatial non-locality and dissipation. It also gives some insight 
into a sufficient condition to observe negative refraction in media with weak non locality.
These results provide a sound basis both for the theoretical investigation of negative refraction in wave physics, and for the design and analysis
of real materials.

%%%\section{The setup}

Maxwell's equations for electrodynamics in material media can be written in full generality in the following form \cite{agranov2006, agranov1984, melrose1991,landau1960, rukhadze1961}:
\begin{align}
\label{eq1}
&\nabla \wedge {\bf B} =  \mu_0 \, {\bf j}_{ext} + \mu_0 \, \partial_t {\bf D} & & \nabla \cdot {\bf D} =  \rho_{ext} \nonumber  \\
&\nabla \wedge {\bf E} =  - \partial_t {\bf B} & & \nabla \cdot {\bf B} = 0
\end{align}
where ${\bf j}_{ext}$ and $\rho_{ext}$ are external current and charge densities and $\mu_0$ is the vacuum permeability. 
In the framework of linear response theory, the field ${\bf D }$ is defined as \cite{agranov1984,melrose1991,landau1960, rukhadze1961}
\begin{equation}
\label{eq2}
{\it D}_i (t,{\bf r}) = \epsilon_0 \int_{-\infty}^{+t} \int_V \epsilon_{ij}(t,t',{\bf r},{\bf r}') E_j (t',{\bf r}') \, dt' d^3r' 
\end{equation}
with $V$ the volume of the medium, $\epsilon_0$ the vacuum permittivity, and where implicit summation over repeated indices is assumed.
 The dielectric function $\epsilon_{ij}(t,t',{\bf r},{\bf r}')$ 
is proportional to the retarded correlator of the current density, and encodes both the electric and magnetic response of the medium \cite{agranov2006,agranov1984,melrose1991, landau1960, rukhadze1961}. In particular, in this formalism, there is no need
to introduce an additional field ${\bf H}$, provided that the full dispersive and non-local behaviour of the dielectric function is taken into account
(see the Appendix). For simplicity, we restrict the analysis to linear, isotropic, non-gyrotropic and homogenous materials, 
and leave generalisations for further studies. Apart from this restriction, all results are derived from first principles, without any 
specification of a particular medium \cite{OBS}. For translation invariant media, one has $\epsilon_{ij}(t,t',{\bf r},{\bf r}') = \epsilon_{ij}(t-t',{\bf r}-{\bf r}')$,
and in Fourier space Eq.\eqref{eq2} reads as
\begin{equation}
\label{eq3}
{\it D}_i (\omega, \vec{k}) = \epsilon_0 \, \epsilon_{ij}(\omega, \vec{k}) E_j (\omega,\vec{k})
\end{equation}
where $\omega$ is the frequency and $\vec{k}$ is the wave vector that describes spatial non-locality. We stress that all
magnetic effects, such as the magnetic response usually described in term of a local permeability $\mu (\omega )$, or even the magnetoelectric activity,
can be encoded into specific spatial non-local properties of the dielectric function \cite{agranov2006, melrose1991,landau1960, rukhadze1961}. A discussion on the relation between this formalism and the usual approach in terms of the fields ${\bf D}$ and ${\bf H}$ is provided
in the Appendix, as well as a justification of the relevance of this formalism for a first principles analysis of the 
electromagnetic response of material media.

The dielectric function $\epsilon_{ij}(\omega,\vec{k})$ has to satisfy various properties~\cite{agranov1984,landau1960, kadano1963, forster1975}. 
Causality implies that $\epsilon_{ij}(t-t',{\bf r}-{\bf r}')$ vanishes for $t<t'$. In the frequency domain, this means that $\epsilon_{ij}(\omega,\vec{k})$ is 
an analytic function of $\omega$ in the positive half-part of the complex plane for which $\hbox{Im }\omega >0$. 
This property leads to the Kramers-Kronig relations
\begin{eqnarray}
\label{eq4}
&&\hbox{Re }  \epsilon_{ij}(\omega, \vec{k})- \delta_{ij} =  \frac{2}{\pi}P \int_{0}^{+\infty} \frac{\omega' \, \hbox{Im } \epsilon_{ij}(\omega', \vec{k}) }{\omega'^2-\omega^2} \, d\omega' \nonumber\\
&&  \hbox{Im } \epsilon_{ij}(\omega, \vec{k}) = - \frac{2 \omega}{\pi}P \int_{0}^{+\infty} \frac{\hbox{Re } \epsilon_{ij}(\omega', \vec{k}) -\delta_{ij} }{\omega'^2-\omega^2} \, d\omega'
\end{eqnarray}
where $P$ stands for principal value.
It is important to highlight that the Kramers-Kronig relations are valid for all values of $\vec{k}$, that essentially plays the role of a parameter in the 
equations, and is, at this stage, independent on $\omega$.
In isotropic media, the dielectric function takes the form
\begin{equation}
\epsilon_{ij}(\omega,k)=\epsilon_{T}(\omega,k) \, \left (\delta_{ij} -  \frac{k_ik_j}{k^2} \right ) + \epsilon_{L}(\omega,k)  \, \frac{k_ik_j}{k^2}
\end{equation}
where $k_i$ are the components of $\vec{k}$ and $k=|\vec{k}|$ \cite{NoteScalar}.
The response of the material is expressed in term of two scalar functions only, the transverse and the longitudinal dielectric functions
$\epsilon_{T}(\omega,k)$ and $\epsilon_{L}(\omega,k)$, that both satisfy the Kramers-Kronig relations~\eqref{eq4}. 

From the Kramers-Kronig relations and the asymptotic behaviour of $\epsilon_{T,L}(\omega,k)$ when $\omega \to \infty$, various sum rules can be derived \cite{forster1975, altarelli1972}, that encode important information and constraints on $\epsilon_{T,L}(\omega,k)$. 
An important sum rule that will be very useful in the analysis is~\cite{altarelli1972}
\begin{equation}
\label{eq5}
 \int_{0}^{+\infty}  \hbox{Im } \epsilon_{T,L}(\omega',k) \, \omega'  \, d\omega' =   \frac{ \pi \omega_p^2}{2} 
\end{equation}
where $\omega_p^2=n e^2/(m \epsilon_0)$ is the square of the plasma frequency, $n$ being the electron density of the material, and $e$ and $m$ the charge and the mass of the electron, respectively. Note that Eq.~\eqref{eq5} is valid for all values of $k$.
Finally, for any passive medium, and for real positive values of $k$ and $\omega$ \cite{OBS2}, the dielectric function satisfies \cite{kadano1963, forster1975}
\begin{equation}
\label{eq6}
\hbox{Im } \epsilon_{T,L} ( \omega, k)   \geq 0 \ .
\end{equation}
Equations (\ref{eq4}), (\ref{eq5}) and (\ref{eq6}) are fundamental relations that constrain the actual electrodynamic properties
of any causal and passive medium.

The frequency $\omega$ and the wave vector $k$ have been treated so far as independent variables, {\it i.e.},
the system is considered "off-shell", as usual in linear response theory \cite{kadano1963, forster1975} (due to the presence of external sources,
 $\omega$ and $k$ do not have to satisfy any relation \cite{agranov1984}). However when considering the propagation of electromagnetic waves
in a medium without external charges or currents, Maxwell's equations impose the dispersion relations
\begin{equation}
\label{eq7}
\epsilon_{T}(\omega,k)= \frac{c^2 k^2}{\omega^2} \qquad,  \qquad \epsilon_{L}(\omega,k)=0
\end{equation}
for the transverse and longitudinal components of the electric field ${\bf E}$, respectively, with $c$ the speed of light in vacuum. 
Equations~\eqref{eq7} define a set of relations between the wave vector $k$ and the frequency $\omega$. 
These relations provide further restrictions to the electrodynamics in the medium. 
In the presence of spatial non-locality, both the transverse and longitudinal components of the electric field can propagate. 
In this Letter we focus on the response to transverse electromagnetic waves.

%%%\section{Transparent media}

We shall first discuss the particular case of transparent media.
In this case, negative refraction means that, in a certain range of frequencies, phase and group velocities
have opposite directions. Indeed, both the electromagnetic energy density and the Poynting vector are well defined, 
the latter being simply the product of the electromagnetic energy density and the group velocity (the group velocity equals
the energy velocity, at least for media with weak spatial non-locality~\cite{Mikki2009}).
For local media, it has already been argued that negative refraction cannot be observed in absence of dissipation~\cite{stockman2006,agranov1984}.
We will now show that the same conclusion holds in the presence of spatial non-locality, namely, that for transverse waves, negative refraction
is not possible in the frequency region of transparency of a medium.

In the transparency region one has $\hbox{Im } \epsilon_{T,L}(\omega,k)=0$, and the Kramers-Kronig relations involve standard integrations.
Deriving Eq.~\eqref{eq4} with respect to frequency~\cite{landau1960,agranov1984}, and using Eq.~\eqref{eq6}, we obtain
\begin{equation}
\label{EQ9}
\frac{d \left [ \hbox{Re } \epsilon_{T,L}(\omega,k) \right ]} { d\omega} \geq 0
\end{equation}
for all frequencies in the transparency region (this is not the case in regions in which absorption cannot be disregarded, where the dielectric
function can satisfy $d [\hbox{Re } \epsilon_{T,L}(\omega,k)]/d\omega < 0$, known as anomalous dispersion).
For transverse waves Eq.~\eqref{eq7} implies
$n^2_T[\omega,k(\omega)]=k^2c^2/\omega^2=\epsilon_{T}[\omega,k(\omega)]$, with $n_T$ the transverse refractive index, �
and hence the dispersion relation $k(\omega)$. The transverse group velocity is defined as
$v_g^T=d \omega(k)/dk= c [d(\omega n_T)/d \omega]^{-1}= c [n_T+\omega d n_T/d \omega]^{-1}$, while the phase velocity is 
$v_p^T=c/n_T$~\cite{vgvp}. Using Eq.~\eqref{EQ9}, noticing that $k$ and $n_T$ are real functions of frequency, we obtain
$n_T  \, d n_T/d\omega > 0$, or equivalently
\begin{equation}
\label{eq10}
v_g^T \, v_p^T> 0 \ .
\end{equation}
This implies that, as a consequence of causality, negative refraction for transverse waves is not possible in the
transparency spectral region of a medium. 

%%%\section{General media}

We shall now perform a general analysis of the conditions for negative refraction, including both dissipation and spatial non-locality. 
For non-transparent media, a consistent definition of concepts such as energy density, energy flux, or heat dissipation rate 
is out of reach (see for example Ref.~\cite{agranov1984}). In particular the group velocity does not define the direction of the energy flow.
 We have to find an alternative way to assess the direction of the energy flow based on the wave vector. 
To proceed, we use Eq.~(\ref{eq7}) to obtain the dispersion relation $k(\omega)$. Since we consider passive media, the direction 
of the energy flux is dictated by the direction of dissipation of energy, or equivalently of the damping of the wave amplitude, the latter
being given by the sign of $\hbox{Im }k$. The direction of the phase velocity is given by $\hbox{Re }k$.
As a result, a general condition for negative refraction is the existence of frequencies $\omega$ such that~\cite{Demis}
\begin{equation}
\label{eq11}
\hbox{Re } [k(\omega)]\cdot \hbox{Im }[k(\omega)] <0 \ ,
\end{equation}
meaning that the damping of the wave amplitude occurs in a direction opposite to the phase velocity.

%%%\subsection{Necessary Condition for Negative Refraction}

We have demonstrated previously that for both local and non-local media, negative refraction is not possible without dissipation. We shall now show that 
spatial non-locality itself is a necessary condition for negative refraction.
Let us consider a dissipative medium for which spatial non-locality is assumed to be negligible. For transverse waves, we have
 $\epsilon_{T}(\omega,k) \equiv  \epsilon^0(\omega)$ and the dispersion relation Eq.~\eqref{eq7} simplifies into
\begin{equation}
\label{eq14}
k(\omega)= \frac{\omega}{c}\sqrt{ \epsilon^0 (\omega)} \ .
\end{equation}
For convenience we choose to represent any complex number $z$ as $z=\rho_z e^{ i \phi_z }$, with $\rho_z\geq0$ and $2 \pi > \phi_z \geq0$.
Due to Eq.~\eqref{eq6}, $\hbox{Im }\epsilon^0 (\omega)  \geq 0$ for real frequencies, and the phase  $\phi_{\epsilon^0}$ 
must satisfy $0 \leq \phi_{\epsilon^0} \leq \pi$. The phase of its square root then satisfies  $0 \leq \phi_{\sqrt{\epsilon^0}} \leq \pi/2$, and hence
\begin{equation}
\label{eq15}
0 \leq \phi_{k}  \leq \pi/2  \ .
\end{equation}
Equation \eqref{eq15} is clearly in contradiction with condition \eqref{eq11}, and we are left with the conclusion that spatial non-locality and 
dissipation are necessary conditions for the existence of negative refraction. This is the first important result in this Letter. Although a 
related discussion can be found in Ref.~\cite{agranov1984}, these necessary conditions have not been stated previously to our knowledge. 
Finally, let us note that the necessity of spatial non-locality also holds for longitudinal waves. 
Indeed, in absence of spatial non-locality the longitudinal component of the 
electric field cannot propagate, and hence cannot exhibit negative refraction. 

%\subsection{Sufficient Condition for Negative Refraction}

We now investigate the existence of a sufficient condition for negative refraction.
For non-gyrotropic media, due to invariance under point reflection, the dielectric functions $\epsilon_{T,L}(\omega,k)$ can be expanded 
in powers of $k^2$ in the form (contributions with odd powers of $k$ are forbidden by symmetry) 
\begin{equation}
\label{eq12}
 \epsilon_{T,L} (\omega,k)= \epsilon^0 (\omega) + \sum_{l=1}^{+\infty} \epsilon^l_{T,L} (\omega) (k \mathcal{L})^{2l}
\end{equation}
where $\mathcal{L}$ denotes the largest intrinsic characteristic length of the material ({\it e.g.} the lattice constant for crystals, the mean free path of conduction 
electrons for metals, the Debye screening length for plasmas, etc), and vanishes in local media.
Inserting Eq.~\eqref{eq12} into the sum rule \eqref{eq5}, and imposing the validity of the resulting equation for any value of $k$, 
we obtain the following set of equations: 
\begin{eqnarray}
\label{eq13}
\int_{0}^{+\infty}  \hbox{Im } \epsilon^0  (\omega') \, \omega' \, d\omega' &=&   \frac{ \pi \omega_p^2}{2}  \nonumber \\
\label{NS}
\int_{0}^{+\infty}  \hbox{Im }  \epsilon_{T,L}^l  (\omega') \, \omega'  \, d\omega' &=&  0,  \qquad  l>0 \ .
\end{eqnarray}
Equations (\ref{eq13}) implies that for $l>0$, the functions $\hbox{Im } \epsilon_{T,L}^l  (\omega)$ have to change sign in the interval 
$\omega \in [0,+\infty]$.
This apparently surprising result is a direct consequence of causality (for further details see the Appendix). Also
note that it is not in contradiction with Eq.~\eqref{eq6} since it only applies separately for each $\epsilon_{T,L}^l  (\omega)$ with $l > 0$, and Im $\epsilon^{0}  (\omega) > 0$, while Eq.~\eqref{eq6} 
implies that the imaginary part of the total sum (\ref{eq12}) (for real $k$) is positive.

For transverse waves, and for isotropic and non-gyrotopic media, we shall now prove that the first non-local correction to the dielectric function
gives a sufficient condition for the existence of negative refraction. Let us consider the dielectric function \cite{OBS3}
\begin{equation}
\label{eq16}
 \epsilon_T (\omega,k) = \epsilon^0 (\omega) + \epsilon_T^1 (\omega) (k\mathcal{L})^2 \ .
\end{equation}
Inserting Eq.~\eqref{eq16} into \eqref{eq7}, we obtain
\begin{equation}
\label{eq17}
k (\omega)=  \frac{\omega}{c} \sqrt{ \epsilon^0 (\omega) \cdot  F (\omega)}
\end{equation}
where $F(\omega) = [1-\omega^2 \mathcal{L}^2 \epsilon_T^1 (\omega)/c^2]^{-1}$.
From Eq.~\eqref{eq13} we have deduced that there exist frequencies $\omega$ such that $\hbox{Im } \epsilon_T^1 (\omega) < 0$,
leading to $\hbox{Im } F(\omega) < 0$. Since $\phi_{k} = \phi_{\sqrt{\epsilon0}} + \phi_{\sqrt{F}}$,
we can state that
\begin{equation}
\label{eq24}
\exists \hbox{  }\omega \hbox{ s.t. } \pi \geq \phi_{k} > \phi_{\sqrt{\epsilon^0}} +  \pi/2 \ .
\end{equation}
This condition on the phase of the vector amounts to showing that frequencies for which Eq.~\eqref{eq11} (defining negative refraction) is satisfied
necessarily exist.
We conclude that a non-local medium, that can be represented as the most general first order correction (allowed by symmetry) to a local response Eq.~\eqref{eq16}, necessarily has a range
of frequencies in which negative refraction is observed. This is the second important result in this Letter. 

% Remarks on the sufficient condition
The existence of the necessary condition based on a generic $k^2$ correction to the local response model deserves two comments.
First, applying the same analysis to longitudinal waves does not allow us to conclude on the existence of a sufficient condition
for negative refraction in this case.
Second, we stress that while dissipation and spatial non-locality are necessary conditions, they are not in general sufficient conditions for negative 
refraction.
% over the full spectral range. 
For example, let us consider a non-local transverse dielectric function of the form
\begin{equation}
\label{eq29}
 \epsilon_T (\omega,k) =  1 - \frac{N_0^2}{ \omega^2 - D_0^2 + i D_1 \omega - D_2 k^2 + i D_3 \omega k^2 } 
\end{equation}
representing the generic response of a spatially non-local (isotropic, homogeneous and passive) medium near a resonance, $N_0$ and $D_j$ being real positive constants.
For particular choices of the parameters, it describes the response of charged viscous fluids ({\it e.g.} electrons in certain metals \cite{forcella2014}), 
or the spatially non-local Lorentz model ({\it e.g.} exciton-polariton resonance for certain semiconductors \cite{Cocoletzi}). 
Changing the values of the coefficients of the $k^2$ terms, one can switch from a negative refraction regime to a positive refraction regime at low frequencies, 
as previously shown for the electrons hydrodynamic model~\cite{forcella2014} (more details and other examples are given in
the Appendix). The analysis therefore leads to the conclusion that spatial non-locality supports, but does not
necessarily imply negative refraction~\cite{note}.

Before concluding it is worth commenting on the relation between the formalism used in this Letter and the more familiar local formalism.
The usual local analysis, in term of the dielectric function $\epsilon(\omega)$ and a magnetic permeability $\mu(\omega)$ is clearly included in the 
non-local formalism described in this Letter in term of a single spatially non-local response function $\epsilon_{ij}(\omega, \vec{k})$. It indeed 
emerges as a particular case in the limit $\omega/k \to 0$ and
$k \to 0$ (see Refs.~\cite{landau1960, rukhadze1961} and the Appendix).
In this limit, our result that spatial non-locality is a necessary condition for negative refraction translates into the well known fact that, 
in the local limit, to observe negative refraction it is necessary to have both $\epsilon(\omega)\neq 1$ and $\mu(\omega)\neq 1$. 
The sufficient condition for negative refraction, usually stated in the local formalism in the form $\hbox{Re } [\epsilon(\omega)] \, |\mu(\omega)|+$  $\hbox{Re } [\mu(\omega)] \, |\epsilon(\omega)| < 0$ (see {\it e.g.} Refs.~\cite{veselago1968, Demis, DL,VeselagoReview}) 
is consistent with (although not implied by) our analysis. Indeed, the sufficient condition of a $k^2$ non-local correction does not require any constraints on
the general form of the non-local permittivity  Eq.~(\ref{eq16}).
However, one should not conclude that the theory described in this Letter is a mere generalisation of a well-established approach. 
The use, in the first place, of a non-local
description of the electrodynamic response is fundamental at least for three reasons. 
First, it is important to note that the meaning of $\mu(\omega)$ as a true response function [with $\hbox{Im }\mu(\omega)\ge 0$ for passive media] is only
valid for very small $k$ {\it and} in a limited range of frequencies in the neighbourhood $\omega/k=0$~\cite{landau1960, rukhadze1961}, see also the Appendix.
Outside this range, $\mu(\omega)$ looses its meaning of a magnetic permeability, and its imaginary part 
has in general an indefinite sign ( see the Appendix). This is actually the reason behind the sufficient condition for negative refraction proven in this Letter [the function $F(\omega)$ in Eq.~\eqref{eq17} can be understood as an effective permeability].
Second, the analysis based on a non-local dielectric function reveals that $\mu(\omega)$, even if apparently a local response, encodes (some of) 
the non-local effects to order $k^2$ initially included in the non-local dielectric function $\epsilon_{T,L} (\omega,k)$ (see the Appendix for a proof).
Therefore, the existence of a permeability $\mu(\omega) \neq 1$, as a condition for negative refraction in the local approach,
 can be seen as a disguised residual of spatial non-locality.
Third, the local formalism, although
appropriate for a phenomenological analysis of certain materials, cannot be used for a first principles analysis of negative refraction on the full spectrum of frequencies, because of its limited range of validity around
$\omega/k = 0$ and $k \to 0$~\cite{melrose1991,landau1960, rukhadze1961}.
In the present Letter we applied constraints coming essentially from causality [except for Eq.~(\ref{eq7})]. In the local limit it is possible to include thermodynamical constraints as well, that are instead in general not valid in the non-local theory. In Ref.~\cite{VMarkel} it has been indeed argued that thermodynamics forbids negative refraction in the local limit. 
The addition of thermodynamical considerations hence strengthen once more the need of non-local approach, and the fact that negative refraction is realised exactly outside the regime of applicability of the $\epsilon(\omega)$ and $\mu(\omega)$
formalism.

In conclusion, we have demonstrated that negative refraction of electromagnetic waves is intrinsically related to both dissipation and spatial non-locality. 
Their simultaneous presence is a necessary condition for negative refraction. The implications of the theory developed in this Letter are manyfold. 
At the fundamental level, the analysis can be extended non-isotropic or non-homogeneous media, or to other kind of waves such as
acoustic waves, for which the design of negative refraction materials is an active field as well. At a practical level, the results in this Letter show
that non locality should not be considered as a refinement of the existing theories, allowing one to improve in an incremental way the precision in
the determination of the effective response functions of real materials. Instead, any inverse procedure should include in the first place non locality,
since its exclusion is fundamentally inconsistent, and can lead to uncontrollable systematic errors. 
Finally, our results open the way to a systematic classification of media with or without negative refraction, providing a clear connection
between this phenomenon and the dynamical properties of materials, as already shown in the case of metals~\cite{forcella2014}.

\begin{acknowledgments}
D. F. would like to thank V. Markel for relevant discussions.
This work was supported by LABEX WIFI (Laboratory of Excellence within the French Program ``Investments for the
Future'') under references ANR-10-LABX-24 and ANR-10-IDEX-0001-02 PSL*.
\end{acknowledgments}

\section{APPENDIX}

%\subsection{Appendix}

\section{A.1 Remarks on the $\epsilon_{ij}(\omega, \vec{k})$ formalism}
In the main text we have chosen a specific formalism to discuss electrodynamics in media. This formalism encodes all the interactions of the medium with the electromagnetic radiation in the field $\bold{D}$ through only the response function $\epsilon_{ij}(\omega,  \vec{k})$ \cite{agranov2006, agranov1984, melrose1991,landau1960, rukhadze1961}, without the need to introduce the macroscopic field $\bold{H}$. In this framework we assess necessary or sufficient conditions for negative refraction. In this short section, building on the existing literature, we briefly explain why this is the appropriate formalism to undergo the first principle analysis we perform in the main text, the relation between this formalism and more well-known formalisms in term of the macroscopic fields $\bold{D}$ and $\bold{H}$, the limit to the local formalism in term of the traditional functions $\epsilon(\omega)$ and $\mu(\omega)$, and we comment on some subtleties.

The formalism for electrodynamics in media based on the response function $\epsilon_{ij}(\omega,  \vec{k})$ is valid for any system \cite{ObsBoundaries} in the linear response regime, without the need of any other specific assumption \cite{melrose1991,landau1960, rukhadze1961}. It encodes, as special case, the usual local formalism in term of the fields $\bold{E}$, $\bold{B}$, $\bold{D}$, $\bold{H}$ and the electric permittivity $\epsilon(\omega )$ and magnetic permeability $\mu (\omega )$, but it is much more general because it includes the spatial non-local effects disregarded in the local formalism \cite{agranov2006,agranov1984,melrose1991, landau1960}. The $\epsilon_{ij}(\omega, \vec{k})$ formalism indeed reduces to the local formalism in the limit of small spatial non-locality and slowly varying fields. The possibility to write Maxwell equations in media in term of a single additional macroscopic field $\bold{D}$, in addition to the the fields $\bold{E}$ and $\bold{B}$, without introducing the field $\bold{H}$ field is indeed quite well known \cite{landau1960}. In the case of spatially homogeneous and isotropic invariant systems the relation between the two formalisms can be easily made explicit \cite{rukhadze1961}. Indeed for spatial non-local media Maxwell equations and the electromagnetic response of the medium can be equivalently written in term of the two vector fields $\bold{D'}$, and $\bold{H}$, and the non-local scalar functions $\epsilon(\omega, k)$ and $\mu(\omega, k)$: $\bold{D'}=\epsilon(\omega, k)$  $\bold{E}$ and $\bold{B}=\mu(\omega,  k)$ $\bold{H}$. These two non-local formalisms are actually completely equivalent \cite{rukhadze1961}. For spatially homogeneous and isotropic media the function $\epsilon_{ij}(\omega,  \vec{k})$ depends only on two scalar functions:  $\epsilon_{T}(\omega,k)$ and  $\epsilon_{L}(\omega,k)$, as explained in the main text.
The formalism in term of $\epsilon_{T,L}(\omega,k)$ and $\bold{D}$, described in the main text, translates the fundamental fact that all the interactions between the electromagnetic field and the matter,  including both electric and magnetic effects, are only due to the electromagnetic current $\bold{j}$. It indeed avoids to separate such effects in term of the two vectors: polarisation and the magnetisation, as instead done by the $\epsilon$ and $\mu$ formalism. Instead $\epsilon_{T,L}(\omega,k)$ are the only fundamental functions: they satisfy all the properties of response functions \cite{kadano1963,forster1975}, they are well defined for all values of $\omega$ and $k$, and they are actually simply real linear functions of the causal correlator of electromagnetic currents: the fundamental observable encoding all the interaction between matter and electromagnetism. Such formalism is more appropriate and easier to handle in the case of time-varying fields. For time-varying fields the usual split of the current in term of polarisation and magnetisation vectors is indeed absolutely arbitrary \cite{rukhadze1961}. The $\epsilon_{T,L}(\omega,k)$ formalism implies to formally fix the magnetisation to zero and define the current only in term of the time derivative of the polarisation vector. The consequence is that the magnetic effects are automatically encoded in the $k$ dependency of the dielectric tensor $\epsilon_{T,L}(\omega,k)$. The relations between the two formalisms goes as follow \cite{rukhadze1961}:
\begin{eqnarray}
\label{eq8}
&&\epsilon(\omega, k) = \epsilon_{L}(\omega,k), \nonumber  \\
&&\mu(\omega, k) =  \Big(1- \frac{\omega^2}{c^2 k^2} \Big( \epsilon_{T}(\omega,k) - \epsilon_{L}(\omega,k)\Big) \Big)^{-1}  \nonumber  \\
\end{eqnarray}
Even if $\epsilon_{T,L}(\omega,k)$ are the most fundamental objects from a theory perspective, it is nevertheless worth to state that the formalism in term of $\epsilon(\omega, k)$ and $\mu(\omega, k)$ results to be more adapted to describe static or almost static fields (that could be more cumbersome in term of $\epsilon_{T,L}(\omega,k)$). Indeed Maxwell equations allow to encode all the effects of a time varying magnetic field $B_T$ in the spatial variation of the transverse component of the  electric field $E_T$, in Fourier:  $B_T=\frac{c k }{\omega} E_T$. From this equation it is however clear that the limit of small spatial non-locality, i.e. $k\rightarrow 0$, and  slowly varying fields, i.e. $\omega\rightarrow 0$, should be performed at the same time to obtain a non zero magnetic field. This makes less straightforward to describe static magnetic fields in term of the $\epsilon_{T,L}(\omega,k)$, compared to the $\epsilon(\omega, k)$ and $\mu(\omega, k)$ formalism.
Even if the formalism in term of  $\epsilon(\omega, k)$ and $\mu(\omega, k)$, could appear more familiar to the reader it has instead some subtleties that are rarely stated in literature, and it is worth discussing here \cite{rukhadze1961}. Clarifying these subtleties would probably make clearer why the $\epsilon_{T,L}(\omega,k)$ formalism should be preferred for a first principles analysis.
As explained in the main text, the causality of the system implies that $\epsilon_{T,L}(\omega,k)$ satisfy the Kramers-Kronig relations, and the associated sum rules. In particular the sum rule for the imaginary part of $\epsilon_{T,L}(\omega,k)$, discussed in the main text, and the second equation in (\ref{eq8}), imply the equation:
\begin{equation}
\label{EQ8}
\int_{0}^{+\infty}  \frac{c^2 k^2}{\omega'} \frac{ \hbox{Im } \mu(\omega', k)}{ \arrowvert \mu(\omega', k) \arrowvert ^2}   d \omega' =0
\end{equation}
This should be valid for any  value of $k$, and it implies that the imaginary part of $\mu(\omega, k)$ has to change sign in the interval $\omega \in  [0;+\infty]$. This is in clear contrast with the case of  $\epsilon_{T,L}(\omega,k)$, whose imaginary part is of definite sign for all frequencies (and real positive $k$), due to the passivity of the medium (see the discussion in the main text). Such apparently wired fact is a direct consequence of the causality of the system.

Both the formalism in term of $\epsilon_{T,L}(\omega,k)$, or the one in term of $\epsilon(\omega,k )$ and  $\mu (\omega,k )$, reduce to the usual local electrodynamics in term of the $k$ independent functions $\epsilon(\omega )$ and  $\mu (\omega )$ in the limit of small spatial dispersion \cite{rukhadze1961, agranov2006,agranov1984}. However such limit can somehow be subtle for what concern the magnetic response. Indeed in the local limit:  $\epsilon_{L}(\omega,0)=\epsilon_{T}(\omega,0)=\epsilon(\omega )$, while it can be shown that the function $\mu(\omega, k)$ assumes the meaning of the magnetic permeability $\mu (\omega )$ only in the narrow frequency band around $\omega/k=0$ for small value ok $k$ \cite{rukhadze1961}. In this limit the resulting function satisfies all the usual properties of a response function \cite{kadano1963, forster1975}, however such properties are in general not valid for the full set of frequencies. In particular it can be shown that the imaginary part of the function obtained taking naively the limit $k\rightarrow0$ for $\mu (\omega,k )$ should not have definite sign. This is once again a consequence of the causality of the system. Nevertheless the resulting function is not interpretable as a magnetic permeability in that range of frequency.

It should hence be clear that a first principles analysis of the response of a generic medium cannot be performed in the local formalism due to two main limitations. The local formalism disregards spatial non-local effects that can become relevant for certain media or for certain range of parameters or frequencies. For example some well-known corrections to the local formalism, due to spatial non-locality, come from the microscopic structure of the medium as in the case of crystals, for short wave length, or due to long range coherent phenomena as for example for plasmas, conducting or hydrodynamic media, where such corrections are relevant also at low frequencies. The other limitation is that the function $\mu (\omega )$ is a response function, and it actually describes the magnetic response of a medium, only for pretty limited set of frequencies \cite{landau1960,rukhadze1961} around $\omega/k=0$, and moreover it disguises in non-transparent way some weak spatial non-localities of the system.
On the other hand in the spatially non-local formalism in term of $\epsilon_{T,L}(\omega,k)$ these functions are well defined response functions for all frequencies and wave vectors. The local formalism results appropriate and better suited only for almost homogeneous and almost constant electric and magnetic fields \cite{melrose1991}. 

\section{A2. Examples of response functions}
In this section we briefly illustrate some explicit examples of spatial non-local electromagnetic responses relevant for phenomenological and experimental applications, and we discuss the effect of spatial non-locality on negative refraction for the associated media. The aim of this analysis is to provide some explicit and concrete examples for the general theoretical analysis performed in the main text. In particular we provide some analysis for the response function introduced in the main text (see also equation (\ref{eq29}) here below) in term of generic resonance, and in particular for the limits in which it describes well-known systems in literature. A general analysis and classification of electromagnetic responses for spatially non-local media is left for future investigations. 

We restrict our analysis to the transverse electric field and we consider the pretty general spatially non-local permittivity introduced in the main text: 
\begin{equation}
\label{eq29}
 \epsilon_T (\omega,k) =  1 - \frac{N_0^2}{ \omega^2 - D_0^2 + i D_1 \omega - D_2 k^2 + i D_3 \omega k^2 } 
\end{equation}
With $N_0$, and $D_j$ real positive constants. As stated in the mail text, it can be easily check that (\ref{eq29}) is a good response function analytic in the complex upper half $\omega$ plane, and moreover its imaginary part is non-negative definite. It represents the response of a spatially non-local medium near a resonance. 
The function $\epsilon_T (\omega,k)$ in (\ref{eq29}) reduces to two physically well understood cases for particular values of the parameters: the response of a charged viscous fluid (like possibly electrons in certain metals \cite{forcella2014}), or the spatially non-local Lorentz model used to describe the electromagnetic response of certain semiconductors near exciton-polariton resonance, as discussed for example in \cite{Cocoletzi}, see below. 
Using $\epsilon_T (\omega,k)$ in (\ref{eq29}) we are able to illustrate some main points of the analysis performed in the main text. In particular we can see explicitly that while spatially non-locality is a necessary condition for negative refraction, it is instead in general not a sufficient condition (contrary to what happens for the special case of a $k^2$ spatial non-local corrections to the local case, as discussed in the main text). Here below we provide examples of models where negative refraction is always present, models for which it is never present, and models where negative refraction appears only for certain values of the parameters of the medium, while it is absent for others. 
The strategy we will use is to start from local models, in which the $k^2$ dependency of the function (\ref{eq29}) is not present or very weak,
and slowly increase the parameters corresponding to spatial non-locality: i.e. $D_2$ and $D_3$ in (\ref{eq29}).
\subsection{The Hydrodynamic case}
For the particular choice of parameters: $N_0 = \omega_p$, $D_1 = 1/\tau$, $D_3 = \eta$, and $D_0 = D_2 =0$, the model (\ref{eq29}) reduces to the response function of a viscous charged fluids in the presence of dissipation \cite{forcella2014}: 
\begin{equation}
\label{eq30}
\epsilon_T (\omega,k) =  1 - \frac{ \omega_p^2}{ \omega^2  + i /\tau \omega + i \eta \omega k^2 } 
\end{equation}
Such function can be used to discuss for example the response to the electromagnetic field of electrons in certain metals. In this case, the electrons' correlation generates an effective viscosity, $\eta$ in equation (\ref{eq30}), that hence induces the related spatial non-locality \cite{forcella2014}, while the lattice of positively charged atoms breaks translation invariance and it hence introduces an effective dissipation of the medium, represented by $1/\tau$ in equation (\ref{eq30}). This encodes the non-conservation of electrons' momentum at microscopic scale, due for example to the presence of a microscopic lattice. $\omega_p$ is the plasma frequency as discussed in the main text. 

\begin{figure}[h]
    \centering
    \includegraphics[width=0.4\textwidth]{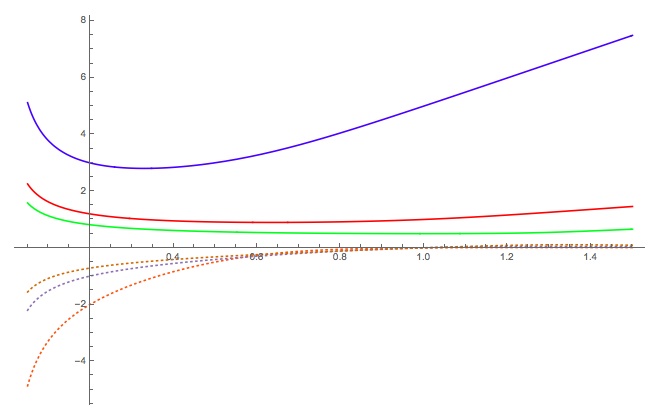}
    \caption{We plot the product Im($k$)  $\cdot$ Re($k$) for the model in equation (\ref{eq30}) with $1/\tau=0$. Continuos lines are associated to the first refractive index, while dashed lines are associated to the second refractive index. The plots are done fixing $\omega_p=1$ and varying the values of the spatially non-local parameter $\eta$. Three values of $\eta$ are represented, respectively $\eta=0.1, 0.5, 1$.}
    \label{FIG1}
\end{figure}

In the particular case of  $1/\tau =0$, i.e. restoration of translation invariance at microscopic scale, the $\epsilon_T (\omega,k)$ in (\ref{eq30}) reduces to the one of charged fluids in absence of impurities or lattice as it can be straightforward derived using Navier-Stokes plus Maxwell equations. In this case it is possible to show that the medium has always two refractive indices, one of which is always negative for frequency below $\omega_p$ \cite{amariti2011}. We plot this result in figure \ref{FIG1}: the refractive index associated to  the dashed lines is indeed always negative.
\subsection{The spatial non-local Lorentz case}
\begin{figure}[h]
    \centering
    \includegraphics[width=0.5\textwidth]{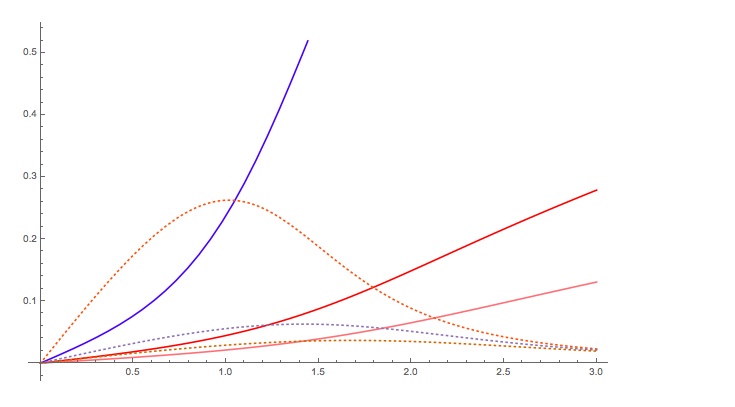}
    \caption{We plot the product Im($k$)  $\cdot$ Re($k$) for the model in equation (\ref{eq31}). Continuos lines are associated to the first refractive index, while dashed lines are associated to the second refractive index. The plots are done fixing $F = 10$,  $\omega_0 = 1$, and $\Gamma = 0.1$ and varying the values of the spatially non-local parameter $D$. Three values of $D$ are represented, respectively $D=0.1, 0.5, 1$. }
    \label{FIG2}
\end{figure}

For the particular choice of parameters: $N_0 = \sqrt{F} \omega_0$, $D_0 = \omega_0$, $D_1 = \Gamma$, $D_2 = D$, and $D_3= 0$, the model ($\ref{eq29}$) reduces to the response function of the Lorentz model modified by spatial non-local effects parametrized by the coefficient $D$:
\begin{equation}
\label{eq31}
 \epsilon_T (\omega,k) =  1 +  \frac{ F \omega_0^2}{ \omega^2_0 - \omega^2  - i \Gamma \omega + D k^2 } 
\end{equation}
As shown in the figure \ref{FIG2} this is a nice example of spatial non-local response with two refractive indices, both positive for all values of the parameters. It is an explicit example showing that spatial non-locality is a necessary, but in general not a sufficient condition for negative refraction.
\subsection{General case}
\begin{figure}[h]
    \centering
    \includegraphics[width=0.5\textwidth]{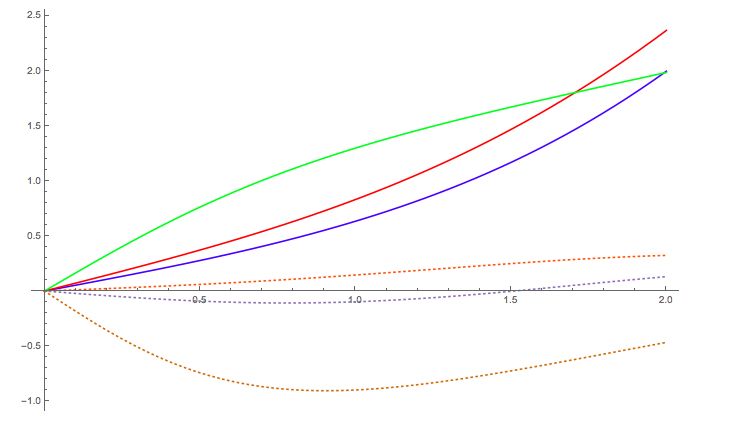}
    \caption{
    We plot the product Im($k$)  $\cdot$ Re($k$) for the model in equation (\ref{eq29}). Continuos lines are associated to the first refractive index, while dashed lines are associated to the second refractive index. The plots are done fixing $N_0 = 5$,  $D_0 = 1$,  $D_1 = 0.8$, and  $D_2 =  0.5 $ and varying the values of the spatially non-local parameter $D_3$. Three values of $D_3$ are represented, respectively $D_3=0.08, 0.15, 0.5$. }
    \label{FIG3}
\end{figure}

In this sub-section we show, as stated in the main text, that the general model represented in equation (\ref{eq29}) exhibits both negative and positive refraction depending on frequencies and the values of parameters. In figure \ref{FIG3} we indeed show that, varying the values of one of the spatial non-local parameters, we observe that there is a threshold of non-locality above which one of the two refractive indices becomes negative. 
A similar phenomenon happens for the hydrodynamic model (\ref{eq30}) with $1/\tau \neq 0$  \cite{forcella2014}, for which one of the two refractive index is negative, below a certain frequency, if $\omega^2 \tau \eta > 1$.


\begin{thebibliography}{99}
\bibitem{OBS} Because we consider homogeneous invariant systems, the theory presented in the paper, and in particular Eq. (\ref{eq3}), describes only the bulk response of a medium, and not its boundaries properties.
\bibitem{vgvp} In isotropic media the group and phase velocities are both along the direction defined by $\vec{k}$. We report only the coefficients of these vectors that define the versus of propagation.
\bibitem{note}
This result can be analyzed using a different approach based on sum rules for the refractive index. It is not described here
for the sake of consistency and brevity, and will be reported elsewhere.
\bibitem{ObsBoundaries} 
Strictly speaking, this is true for spatially homogenous systems. Namely $\epsilon_{ij}(\omega, \vec{k})$ is able to describe the response of a generic medium in the bulk, while the presence of boundaries requires additional informations, such as for example: additional boundary conditions, or the introduction of additional information coming from a microscopic analysis of the medium. When a macroscopic approach is possible the analysis should in general be performed in term of the response function $\epsilon_{ij}(t,t',\bold{r}, \bold{r'})$ in the direct space.
\end{thebibliography}
\end{document}